\documentclass[fleqn,usenatbib]{rasti}

\usepackage{newtxtext,newtxmath}

\usepackage[T1]{fontenc}

\DeclareRobustCommand{\VAN}[3]{#2}
\let\VANthebibliography\thebibliography
\def\thebibliography{\DeclareRobustCommand{\VAN}[3]{##3}\VANthebibliography}

\usepackage{graphicx}	
\usepackage{amsmath}	

\title[XMST: Objective Fracture-Scale Selection]{XMST: An Extended Minimum Spanning Tree Framework with Objective Fracture-Scale Selection}

\author[Mark J. Gallaway]{
Mark J. Gallaway \thanks{E-mail: m.j.gallaway.astro@gmail.com}
\\
}
\usepackage{placeins}
\date{Accepted XXX. Received YYY; in original form ZZZ}

\pubyear{\the\year{}}

\begin{document}
\label{firstpage}
\pagerange{\pageref{firstpage}--\pageref{lastpage}}
\maketitle

\begin{abstract}
We present XMST, an extended minimum spanning tree framework for identifying spatially coherent stellar structures using an objective, data-driven fracture-scale criterion. XMST combines percolation analysis with Jenks Natural Breaks optimisation, restricting the MST edge-length distribution to the subcritical regime before determining the final fracture scale. The method was validated using 1000 controlled Monte Carlo realisations containing eight empirical stellar-association templates, including a deliberately overlapping pair. The Percolation–Jenks fracture scale was highly reproducible, with a median of 17.028 pc and a between-realisation standard deviation of 0.066 pc. The injected structures were recovered with a mean completeness of 0.9996 and mean purity of 0.7413, while alternative CDF and Mean Edge criteria selected substantially larger fracture scales and produced markedly lower purities.
Propagation of the published distance uncertainties through 10,000 additional XMST reconstructions produced a median absolute fracture-scale change of only 0.282 pc. Under these perturbations, mean completeness was 0.8605 and mean purity 0.7126, while 99.30 per cent of injected-template cases continued to satisfy the adopted detection criterion. Individual membership was less stable, with peripheral and locally sparse members showing lower persistence than structural cores. The deliberately overlapping pair was resolved spatially in only 15 of 1000 realisations; a reddening-based post-processing diagnostic increased this to 446, although with a corresponding completeness–purity trade-off. XMST therefore provides a robust method for identifying candidate stellar structures and their stable cores, while detailed outer membership and strongly overlapping populations require additional astrophysical information.
\end{abstract}

\begin{keywords}
methods: statistical -- methods: data analysis -- open clusters and associations: general -- stars: massive -- astrometry -- Galaxy: structure
\end{keywords}



\section{Introduction}

OB associations are loose groups of young, hot O- and B-type stars found predominantly within the spiral arms of the Milky Way and other spiral galaxies \citep{Ambartsumian1947}. Because these luminous stars are relatively short-lived, OB associations trace recent star formation and can be used to delineate the spiral structure of their host galaxies. 
OB associations can span tens to hundreds of parsecs, contain nested subgroups and filaments, and overlap with neighbouring young stellar populations \citep{Blaauw1964}. Consequently, association membership is not always straightforward to determine from observational data alone and depends on the available astrometry, the parameter space being considered, the background population, and the clustering method used to distinguish associations from field stars and overlapping young populations (e.g. Wright \& Mamajek 2018).

The Hipparcos census of nearby associations demonstrated that astrometric membership could be determined systematically using parallaxes and proper motions, while also illustrating the limitations imposed by distance, background contamination, and unfavourable kinematics \citep{deZeeuw1999}. However, the precision of Hipparcos parallaxes increasingly limited reliable membership determination at greater distances, making the identification of more distant association members particularly difficult \citep{Perryman1997}.

Gaia has transformed this problem by providing high-precision astrometry for vastly larger stellar samples. These data have shown that OB associations are frequently more complex than the classical picture of a single, smooth moving group or the expanded remnant of a compact cluster. This is consistent with models in which giant molecular clouds undergo hierarchical fragmentation during star formation, producing structure over a range of spatial scales \citep{Hopkins2013}. Observationally, smaller stellar groups and associations may therefore form part of much larger coherent structures \citep{Wright2018}.

A number of statistical approaches have been developed to identify stellar structures, including OB associations. Density-based methods identify clusters as regions of enhanced local density separated from the surrounding background population. A widely used example is Density-Based Spatial Clustering of Applications with Noise \citep{Ester1996}, which uses a neighbourhood search radius, $\epsilon$, and a minimum number of neighbouring points, minPts, to define clustered regions. However, because the neighbourhood scale is fixed, standard DBSCAN can struggle when applied to datasets containing structures with different spatial densities \citep{Ankerst1999}. The choice of an appropriate scale therefore remains an important part of the clustering problem. 

OPTICS \citep{Ankerst1999}  extends this density-based approach by producing a hierarchical representation of the clustering structure over a range of density scales. This makes it better suited than standard DBSCAN to datasets containing structures with varying densities. However, extracting individual clusters from the resulting hierarchy still requires decisions about how that structure is interpreted.

Minimum spanning trees (MSTs) are a graph-theoretic technique \citep{Kruskal1956} that has subsequently been widely used for cluster identification in astronomy and other scientific disciplines. Each data point is represented as a node within a chosen parameter space, with the separation between neighbouring nodes represented by the length of the connecting edge. The minimum spanning tree is the network that connects all nodes while minimising the total edge length and containing no closed loops. Because the MST is determined by the relative positions of the data points rather than by an assumed cluster geometry, it is particularly useful for identifying irregular or extended structures \citep{Gallaway2012}.

The MST is subsequently fractured at a characteristic edge length, referred to here as the fracture scale, producing a set of disconnected subtrees. Subtrees containing fewer than a specified minimum number of members, together with isolated orphan nodes, are discarded. The remaining subtrees are retained as candidate structures within the dataset. This approach allows irregular, elongated and hierarchical stellar structures to be identified without assuming a particular cluster geometry. 

A major difficulty with classical MST methods is determining an appropriate fracture scale, particularly where clustering occurs over multiple spatial scales. \cite{Gutermuth2009} addressed this by constructing the cumulative distribution function (CDF) of MST edge lengths and identifying a transition between the shorter edges associated with clustered structure and the longer edges linking separate structures. As discussed by Gallaway (2012), however, this transition can
be poorly defined, making the resulting fracture scale difficult to determine consistently.
Hierarchical Density-Based Spatial Clustering of Applications with Noise \citep{Campello2015}
extends density-based clustering by constructing a hierarchy over a range of density scales and selecting clusters according to their stability. HDBSCAN uses mutual-reachability distances to construct a minimum spanning tree from which the cluster hierarchy is derived. It therefore avoids imposing a single global neighbourhood scale and is well suited to datasets containing background noise and structures with varying densities.

HDBSCAN has already been applied successfully to large samples of young and massive stars identified with Gaia. \cite{Chemel2022}  applied the method to approximately 47,700 OB stars and young-cluster tracers from Gaia EDR3, while \cite{Quintana2026}  applied HDBSCAN to a sample of approximately 25,000 O- and B-type stars within 1 kpc and identified 56 high-confidence associations. These studies demonstrate the effectiveness of hierarchical density-based clustering for identifying young stellar structures in large astrometric datasets. 

Existing methods therefore have complementary strengths and limitations. Density-based approaches can accommodate structures over multiple spatial scales but remain dependent on their adopted density definitions and parameter choices, while classical MST methods preserve the underlying geometric structure but require the selection of a fracture scale. In addition, observational uncertainties are not always propagated through the clustering procedure when assessing the stability of recovered memberships.

XMST extends the classical MST workflow by incorporating objective fracture-scale selection, Monte Carlo uncertainty propagation and quantitative stability statistics. Repeated reconstruction of the MST using perturbed realisations of the data allows the stability of both the derived fracture scale and the recovered structures to be measured directly.

XMST addresses the fracture-scale problem using a two-stage Percolation–Jenks procedure. Percolation analysis first identifies the transition from locally connected structures towards the formation of a field-spanning network, defining the upper edge-length range considered. Jenks Natural Breaks optimisation is then applied to the remaining subcritical edge-length distribution to determine the final fracture scale. This provides a reproducible, data-driven alternative to the traditional CDF and Mean Edge fracture criteria.

While HDBSCAN modifies the distance metric to produce a density-aware hierarchy, XMST retains the Euclidean spatial MST and augments its segmentation with objective fracture-scale selection, uncertainty propagation and post-detection stability analysis. XMST is therefore intended as a complementary approach to HDBSCAN and other clustering methodologies, providing an independent means of identifying and assessing candidate stellar structures.

In this paper we describe the XMST methodology and assess its performance using controlled Monte Carlo datasets. We test the stability and reproducibility of the Percolation–Jenks fracture scale, quantify the recovery of injected stellar associations, compare its performance with the traditional CDF and Mean Edge fracture criteria, and investigate the use of reddening as an additional diagnostic for overlapping stellar populations. Application of XMST to the published OB-association catalogue, together with a detailed comparison with existing clustering methodologies, will be presented in a subsequent paper.

\section{Method}

XMST uses a multi-stage procedure to identify candidate stellar associations and quantify the robustness of their recovered memberships. The procedure consists of coordinate transformation, construction of the global minimum spanning tree, determination of the fracture scale using the Percolation–Jenks procedure, extraction of candidate subtrees, and subsequent stability analysis.

\subsection{Input Data}
The parent stellar sample used in this work was taken from the catalogue of candidate O- and B-type stars compiled by \cite{Quintana2025}, hereafter Q25. This catalogue contains 24,706 candidate OB stars within 1 kpc of the Sun and was constructed using astrometric and photometric measurements including data from Gaia Data Release 3. We did not perform an independent selection of OB-star candidates from the full Gaia DR3 archive.

The stellar sample used in this work therefore follows the selection criteria and quality cuts described by Q25, with no additional astrophysical selection applied prior to the XMST analysis. The core XMST algorithm requires only a unique source identifier and three-dimensional positional information. The catalogue heliocentric Cartesian coordinates were therefore retained for construction of the spatial graph, as described in Section 2.2.

Additional parameters were retained only for validation and post-processing. Photometric measurements were used for the reddening experiments described in Section 2.9. No previously published association memberships were supplied to XMST during cluster identification.

\subsection{Coordinate Transformation}
The input catalogue provides heliocentric Cartesian positions for each star derived from Galactic longitude, Galactic latitude and distance. These coordinates were used directly for construction of the spatial graph. They are equivalent to the transformation:

\begin{align}
X &= d \cos b \cos l \\
Y &= d \cos b \sin l \\
Z &= d \sin b
\end{align}

where l and b are Galactic longitude and latitude, respectively, and d is the stellar distance. The Sun is located at the origin of this coordinate system.
Three-dimensional Euclidean separations between stars were calculated from these Cartesian coordinates and used throughout the XMST analysis. This provides the spatial metric used for construction of the Delaunay graph and the subsequent minimum spanning tree.
For the distance-uncertainty realisations described in Section 2.6, the direction of each star was held fixed while its radial distance was perturbed. The Cartesian position was then recalculated from the perturbed distance before reconstruction of the Delaunay graph and minimum spanning tree.

\subsection{Construction of the Global MST}
Following the coordinate transformation, the complete stellar sample is represented as a three-dimensional point cloud in Cartesian space. A global minimum spanning tree is then constructed connecting all stars within the sample. 

\subsubsection{Delaunay Graph}
A naïve implementation would require consideration of all possible pairs of stars, producing a fully connected graph containing $N(N-1)/2$ edges. For the large catalogues considered here, this would be computationally expensive in both memory usage and processing time. Instead, XMST first constructs a three-dimensional Delaunay triangulation. The resulting Delaunay graph forms a sparse network that contains the Euclidean minimum spanning tree \citep{deBerg2008}, allowing the exact global MST to be recovered without considering every possible stellar pair.

\subsubsection{Minimum Spanning Trees}
As only spatial information is used during graph construction, no additional weighting was applied to the Delaunay edges. Edge weights were therefore defined solely by the three-dimensional Euclidean separation between neighbouring stars.

The Delaunay graph was then used as the input to Kruskal's algorithm to construct the global minimum spanning tree. Kruskal's algorithm \citep{Kruskal1956} iteratively selects the shortest available edge that does not create a closed loop, continuing until all stars are connected within a single spanning tree. The resulting global MST forms the basis for all subsequent percolation, fracturing and cluster-identification stages of the XMST algorithm. 

\subsection{Determining the Fracture Scale}
In order to extract candidate structures from the global MST, a characteristic edge length must be identified that separates the shorter edges predominantly connecting stars within local structures from the longer edges connecting those structures to each other and to the surrounding field. This value is referred to here as the fracture scale. 

\subsubsection{Traditional CDF Method}
The traditional cumulative distribution function (CDF) method was also evaluated as a means of determining the fracture scale. Following the approach used in earlier MST studies, the empirical CDF of the complete global MST edge-length distribution was represented by two linear regimes. Least-squares straight-line fits were made to the shorter- and longer-edge sections of the CDF, with the division between the two regimes selected by minimising the combined residuals of the two fits. The intersection of the fitted lines was then adopted as the CDF fracture scale.

The CDF and Mean Edge comparisons are intended specifically to test the effect of fracture-scale selection while retaining the same underlying MST and recovery procedure. A direct comparison with HDBSCAN would additionally introduce differences in the distance metric, hierarchical construction and cluster-selection procedure, and is therefore reserved for the subsequent application paper.

Where the fitted lines did not produce a valid intersection within the observed edge-length range, the midpoint of the two edge lengths surrounding the optimum division was used instead. As discussed by \cite{Gallaway2012}  and  \cite{Kirk2014}, this type of CDF transition can be poorly constrained in complex edge-length distributions. The performance of the CDF fracture criterion was therefore tested directly against the Percolation–Jenks method using the controlled Monte Carlo datasets described in Section 2.8, with the results presented in Section 3.4.

A second simple comparison criterion was also tested using the mean length of all edges in the complete global MST. This value, referred to here as the Mean Edge fracture scale, was applied directly as the threshold for fracturing the tree. Unlike the Percolation–Jenks procedure, no attempt was made to restrict the edge-length distribution or identify separate edge-length populations before calculating the threshold. The Mean Edge criterion was evaluated using the same Monte Carlo realisations as the CDF and Percolation–Jenks methods, allowing their resulting fracture scales and recovery statistics to be compared directly.

\subsubsection{Percolation Analysis}
Following construction of the global minimum spanning tree, the MST edges were sorted into ascending order of length. The graph was then reconstructed by progressively adding edges from shortest to longest. After each edge was added, the connected-component structure was recalculated and the size of the largest connected component recorded.
Initially, the graph consists of small, locally connected groups. As progressively longer edges are introduced, these groups merge and the size of the largest connected component increases. The critical transition edge was defined as the edge producing the largest single increase in the membership of the largest connected component.

If $l_i$ is the length of this critical edge and $l_{i-1}$ is the length of the immediately preceding edge, the operational percolation limit was defined as:

\begin{equation}
L_{\mathrm{perc}} = \frac{l_{i-1} + l_i}{2}
\end{equation}

This places the operational limit immediately below the edge responsible for the largest increase in global connectivity. Only MST edges with lengths less than or equal to this limit were retained for the subsequent Jenks Natural Breaks optimisation.

\subsubsection{Jenks Optimisation}
The percolation analysis identifies the upper edge-length range over which the global MST changes from locally connected structures towards a more globally connected network. To determine the final fracture scale, two-class Jenks Natural Breaks optimisation was applied to the distribution of MST edge lengths less than or equal to the percolation limit \citep{Jenks1967}.

Jenks optimisation identifies the class boundary that minimises the combined within-class variance of the two resulting edge-length populations. In this application, the shorter-edge population is interpreted as being dominated by connections within locally coherent stellar structures, while the longer-edge population is dominated by connections between those structures. The boundary between the two Jenks classes was adopted as the final XMST fracture scale. 

\subsubsection{Bootstrap validation of the fracture scale}
The within-realisation uncertainty in the final fracture scale was assessed using bootstrap resampling \citep{Efron1979}. After the percolation limit had been determined, the subcritical MST edge-length distribution containing edges less than or equal to this limit was resampled with replacement. Jenks Natural Breaks optimisation was then repeated for each bootstrap sample. A total of 200 bootstrap realisations was generated for each Monte Carlo field, and the standard deviation of the resulting Jenks fracture scales was adopted as the within-realisation fracture-scale uncertainty.

\subsection{Fracturing the Tree}
The global MST was fractured at the scale determined by the Percolation–Jenks procedure. All edges longer than the adopted fracture scale were removed, dividing the global MST into a set of disconnected subtrees. Isolated nodes were discarded, together with subtrees containing fewer than the adopted minimum membership of 10 stars. The remaining subtrees were retained as candidate stellar associations. 

\subsection{Stability Analysis}
\label{sec:stability}

To assess the stability of the recovered structures against the published
distance uncertainties, repeated realisations of the stellar catalogue were
generated by perturbing the distance of each star using the uncertainty
intervals reported by Q25. These provide a median distance, $d_{50}$,
together with the 16th- and 84th-percentile limits, $d_{16}$ and $d_{84}$,
respectively.

As the full posterior distributions are not publicly available, these values
were used to construct an asymmetric approximation to the distance
distribution for each star. Distances were sampled using a percentile-matched
two-sided normal distribution centred on $d_{50}$. Separate scale parameters
were used below and above the median so that the quoted 16th and 84th
percentiles were reproduced. The lower and upper scales were defined as:

\begin{equation}
\sigma_{-} =
\frac{d_{50} - d_{16}}
{\left|\Phi^{-1}(0.16)\right|}
\label{eq:sigma_lower}
\end{equation}

\begin{equation}
\sigma_{+} =
\frac{d_{84} - d_{50}}
{\Phi^{-1}(0.84)}
\label{eq:sigma_upper}
\end{equation}

where $\Phi^{-1}$ is the inverse cumulative distribution function of the
standard normal distribution.

A random standard-normal deviate, $z$, was generated for each star. For
$z < 0$, the perturbed distance was calculated using $\sigma_{-}$, while for
$z \geq 0$, $\sigma_{+}$ was used. Any draw producing a non-positive distance
was rejected and resampled.

For stars repositioned during construction of the Monte Carlo datasets, the
fractional lower and upper distance uncertainties of the original Q25
catalogue star were preserved and applied to its new distance. The
repositioned distance was treated as the median distance, with corresponding
16th- and 84th-percentile limits calculated from the original fractional
uncertainties. This retained the asymmetric form and relative scale of the
published uncertainty while allowing the injected structure to be placed at
a different distance within the simulated field.

For each uncertainty realisation, the direction of each star from the Sun was
held fixed while its radial distance was perturbed. The Cartesian position
was then recalculated using the sampled distance. A new Delaunay graph and
global minimum spanning tree were constructed, and the Percolation--Jenks
procedure was repeated independently to determine a new fracture scale. The
resulting MST was then fractured and the recovered subtrees recorded.

Recovered subtrees were compared with those obtained from the corresponding
unperturbed realisation using the Jaccard similarity index. For each nominal
recovered subtree, the perturbed subtree with the highest Jaccard similarity
was selected as its corresponding match. Each star was then assigned a
membership persistence, defined as the fraction of uncertainty realisations
in which it remained associated with the corresponding best-matching
recovered subtree. This provides a measure of the stability of individual
stellar assignments under the published distance uncertainties.

To investigate whether membership persistence depended on position within an
injected structure, three complementary spatial measures were examined.
First, for each injected template, a robust Cartesian centre was defined from
the component-wise median of the stellar $X$, $Y$ and $Z$ positions. Members
were then ranked within their template by their three-dimensional Euclidean
distance from this centre. Second, local stellar density was characterised
using the distance to the fifth-nearest template member, with larger distances
corresponding to locally sparser regions. Finally, a three-dimensional convex
hull was constructed for each injected structure, allowing stars on the hull
boundary to be compared with those lying within its interior. Spearman rank
correlations were used to quantify the relationship between membership
persistence and both radial position and local density.

\subsection{Cluster Comparison Metrics}
To assess the performance of XMST, the recovered subtrees were compared with the known injected structures using a number of complementary metrics. No single statistic fully describes the quality of a recovered structure, since a group may contain most of the genuine members while also including substantial contamination. We therefore considered membership similarity, completeness, purity and recovery when assessing the resulting subtrees.

\subsubsection{Jaccard Index}

The Jaccard Index measures the similarity between two sets of stellar memberships by comparing the number of stars common to both with the total number of unique stars contained within either set. For two groups A and B, the Jaccard Index is defined as:

\begin{equation}
    J(A, B) = \frac{|A \cap B|}{|A \cup B|}
\end{equation}

A value of 1 indicates identical memberships, while a value of 0 indicates that the two groups have no members in common.
In the distance-uncertainty analysis, the Jaccard Index was used to identify the best-matching recovered subtree between each perturbed realisation and the corresponding structure in the unperturbed dataset. This allowed changes in group membership under the published distance uncertainties to be quantified independently of the injected-template recovery statistics.

\subsubsection{Recovery}
The recovered XMST subtrees were compared with the known injected structures in each Monte Carlo dataset to quantify the performance of the algorithm. For each injected structure, the recovered subtree containing the greatest number of its injected members was selected as the best match.

More than one injected structure was permitted to select the same recovered subtree as its best match. The detection statistic therefore measures whether the membership of an injected structure was recovered and does not by itself imply that neighbouring injected structures were resolved as separate subtrees. Separation of the deliberately overlapping pair was assessed independently.

An injected structure was considered detected when its best-matching subtree contained at least 50 per cent of the injected members, subject to the adopted minimum retained group size of 10 stars. A stricter recovery criterion was also defined, requiring a completeness of at least 0.8 and a purity of at least 0.5.

Completeness was calculated as:
\begin{equation}
\mathrm{Completeness} = \frac{N_{\mathrm{correct}}}{N_{\mathrm{injected}}}
\end{equation}
where $N_{\mathrm{correct}}$ is the number of injected members contained within the best-matching recovered subtree and $N_{\mathrm{injected}}$ is the total number of members in the injected structure.

Purity was calculated as:
\begin{equation}
\mathrm{Purity} = \frac{N_{\mathrm{correct}}}{N_{\mathrm{recovered}}}
\end{equation}
where $N_{\mathrm{recovered}}$ is the total membership of the best-matching recovered subtree, including both genuine injected members and contaminating stars.

\subsection{Monte Carlo Validation Design}

The performance of XMST was assessed using controlled Monte Carlo datasets constructed from the parent OB-star catalogue described in Section 2.1. Eight empirical stellar structures were used as injected templates. Their internal three-dimensional spatial distributions were retained from the catalogue, preserving the observed morphology of each template rather than generating idealised spherical or Gaussian clusters. These templates were selected from structures recovered by the nominal XMST analysis of the Q25 parent catalogue and were used to provide realistic, non-parametric stellar morphologies for the controlled simulations.

The eight templates contained 192, 183, 109, 57, 51, 43, 37 and 25 stars, respectively, giving a total template population of 697 stars. These stars were removed from the parent catalogue before construction of the Monte Carlo fields, leaving 24,009 non-template catalogue stars as a fixed comparison population. For each Monte Carlo realisation, the eight templates were independently rotated and repositioned within this population. The transformed templates were then reinserted into the comparison population, giving a total of 24,706 stars in every Monte Carlo realisation. This preserved the internal geometry of each structure while changing its position and orientation relative to the surrounding catalogue stars.

Two of the largest templates were deliberately placed along approximately the same line of sight, with a radial separation selected between 57 and 83 pc. This produced a controlled spatial-overlap case in which the two structures remained physically separated in distance but were difficult to distinguish using spatial clustering alone. The remaining six templates were positioned independently within the comparison field, with their centres restricted to heliocentric distances of 250–900 pc and required to be separated from previously placed template centres by at least 120 pc.

An initial set of 100 Monte Carlo realisations was generated and analysed. This was subsequently extended to 1000 realisations to determine whether the fracture-scale distribution and cluster-recovery statistics had converged. For every realisation, a new three-dimensional Delaunay graph and global minimum spanning tree were constructed and an independent Percolation–Jenks fracture scale was determined. The resulting subtrees were then compared with the known injected memberships using the metrics described in Section 2.7.

For comparison, the same 1000 Monte Carlo realisations were reanalysed using the CDF and Mean Edge fracture criteria in place of Percolation–Jenks. All other aspects of the MST construction, subtree extraction and recovery analysis were kept unchanged. This allowed the effect of the fracture-scale selection procedure itself to be tested using identical stellar fields and injected structures. The CDF and Mean Edge comparisons are therefore intended specifically to assess alternative methods of fracturing the same underlying Euclidean MST. A direct comparison with HDBSCAN would additionally introduce differences in the distance metric, hierarchical construction and cluster-selection procedure, and is therefore reserved for the subsequent application paper.

The effect of the published distance uncertainties was assessed separately using the first 100 Monte Carlo fields. Each field was subjected to 100 independent distance-uncertainty realisations, giving a total of 10,000 perturbed spatial reconstructions. For each realisation, the distance of every star was resampled using the asymmetric uncertainty treatment described in Section 2.6. The Cartesian coordinates, Delaunay graph, global MST and Percolation–Jenks fracture scale were then recalculated independently before the resulting subtrees were recorded and compared with those from the corresponding unperturbed field. No other observational parameters were perturbed during this experiment.

Additional sensitivity tests were performed to determine whether the reported recovery statistics depended strongly on the adopted minimum retained group size or on the thresholds used to define recovery. The minimum group size was tested at Nmin = 5, 10, 15 and 20. The recovered memberships were also re-evaluated using detection thresholds between 0.4 and 0.7 of the injected membership, strict completeness thresholds between 0.7 and 0.9, and strict purity thresholds between 0.4 and 0.7. These tests were used to determine whether the principal recovery results depended critically on the particular threshold values adopted for the main analysis.

\begin{figure*}
    \centering
    \includegraphics[width=\textwidth]{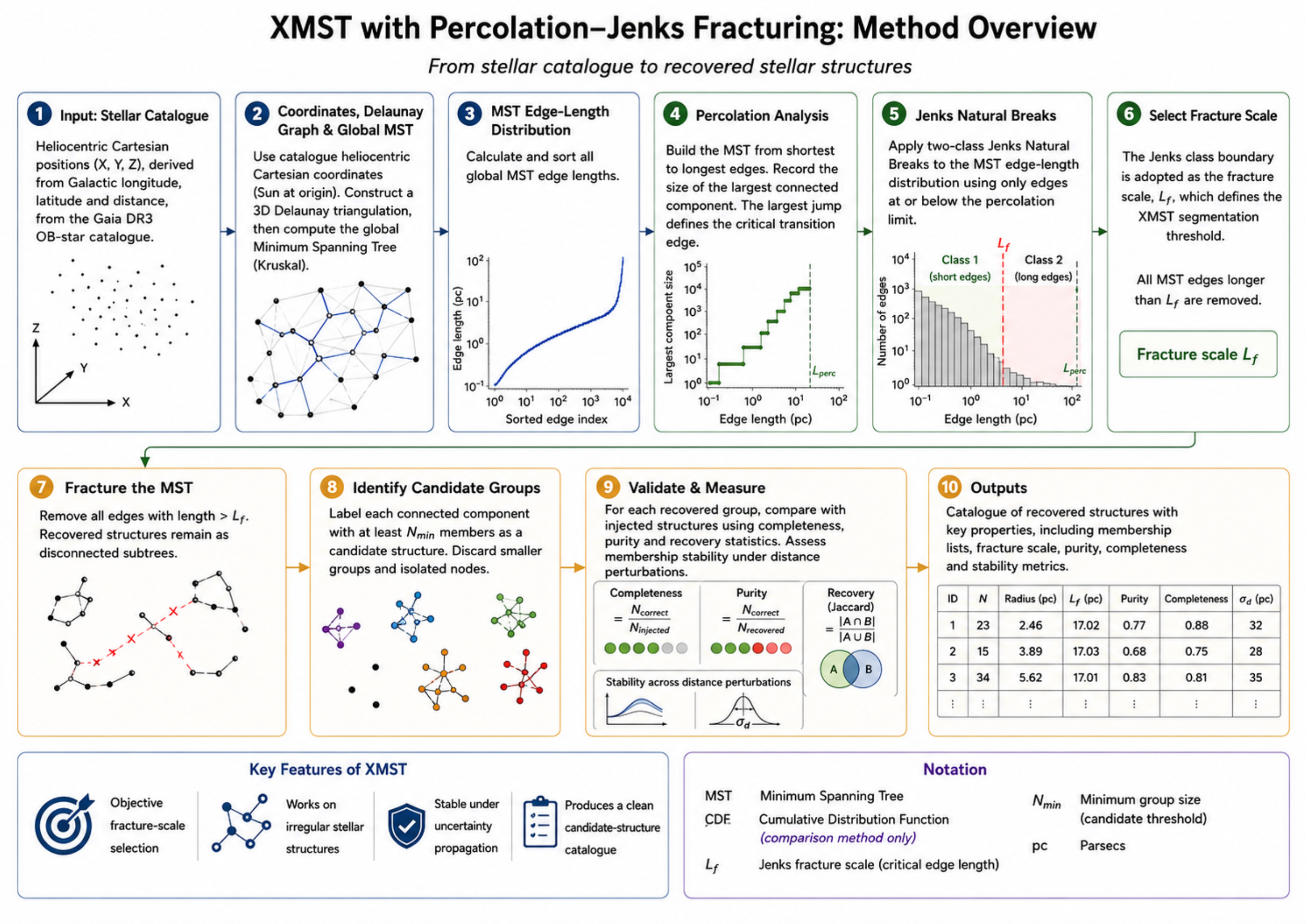}
    \caption{Flow chart of the XMST analysis pipeline, from construction of the global MST to fracture-scale selection and recovery of candidate stellar structures.}
    \label{fig:xmst_flow}
\end{figure*}

\subsection{Reddening Tests}
Initial experiments investigated the inclusion of reddening directly as an additional dimension within the XMST clustering space. This did not produce realistic improvements in the recovery of the injected structures and the approach was therefore abandoned. Reddening was instead retained as a post-processing diagnostic to test whether a spatially recovered structure might contain two populations overlapping along the line of sight.

For each recovered subtree, possible divisions along the existing MST branches were examined. Candidate splits were required to produce two branches containing at least 20 stars each. Each branch was also required to contain $A_V$ measurements for at least 15 stars and for at least 60 per cent of its total membership.

For each branch, the central reddening was represented by the median $A_V$, and the scatter was estimated initially using the median absolute deviation scaled to the Gaussian standard-deviation convention:

\begin{equation}
    s = 1.4826 \times \text{median}\left(|A_{V,i} - \text{median}(A_V)|\right)
\end{equation}

Where this produced a zero or non-finite scatter, the interquartile range divided by 1.349 was used instead. If this also failed to provide a positive finite estimate, the sample standard deviation was used.

The scatter of the two branches was combined with equal weighting:

\begin{equation}
    s_{\text{pooled}} = \sqrt{\frac{s_A^2 + s_B^2}{2}}
\end{equation}

A robust reddening effect size was then defined as:

\begin{equation}
    D_{\text{robust}} = \frac{|\text{median}(A_{V,A}) - \text{median}(A_{V,B})|}{s_{\text{pooled}}}
\end{equation}

Candidate divisions were required to have $D_{\text{robust}} \ge 1.5$. The two reddening distributions were also compared using a two-sided Mann--Whitney $U$ test \citep{Mann1947}. As multiple possible divisions were tested within each subtree, the resulting probabilities were corrected using the Benjamini--Hochberg false-discovery-rate procedure \citep{Benjamini1995}, adopting $q \le 0.01$.

Candidate divisions satisfying both the effect-size and corrected-significance criteria were ranked using the robust effect size and the relative balance of the two resulting branches. The population-balance term was defined as:

\begin{equation}
B = \frac{\min(N_A, N_B)}{\max(N_A, N_B)}
\end{equation}

The preliminary ranking statistic was then defined as:
\begin{equation}
R = D_{\mathrm{robust}} \times B
\end{equation}

To limit the computational cost of the subsequent bootstrap analysis, a maximum of the 25 highest-ranked candidate divisions within each parent structure were retained for further testing.

The statistical stability of each retained candidate was assessed using 500 bootstrap resamples \citep{Efron1979}. The available $A_V$ values were resampled independently with replacement within each branch while preserving the original sample sizes. For each resample, the robust effect size and the direction of the median reddening difference were recalculated. Bootstrap stability, $f_{\mathrm{boot}}$, was defined as the fraction of resamples retaining both $D_{\mathrm{robust}} \ge 1.5$ and the original direction of the reddening difference. A candidate division was required to have $f_{\mathrm{boot}} \ge 0.95$.

A spatial-persistence requirement was also applied to prevent a strong reddening difference from producing a division between branches that were not themselves spatially coherent. For each branch, the spatial edge threshold was progressively reduced and the lowest threshold at which the largest connected descendant retained at least 80 per cent of the original branch membership was determined. The higher of the two branch thresholds was adopted as the common coherence threshold, $L_{\mathrm{coherent}}$. Topology persistence was then defined relative to the global XMST fracture scale, $L_{\mathrm{fracture}}$, as:

\begin{equation}
P = \max\left[ 0, \frac{L_{\mathrm{fracture}} - L_{\mathrm{coherent}}}{L_{\mathrm{fracture}}} \right]
\end{equation}

Candidate divisions were required to have $P \ge 0.10$.

Where more than one candidate satisfied the effect-size, significance, bootstrap-stability and topology-persistence criteria, the preferred division was selected using the combined score:

\begin{equation}
S = D_{\mathrm{robust}} \times B \times f_{\mathrm{boot}} \times P
\end{equation}

The highest-scoring accepted split was applied to the parent structure and the procedure repeated recursively on the resulting branches, to a maximum recursion depth of three.

Gaussian-mixture diagnostics were calculated as supporting evidence following an accepted division. One- and two-component Gaussian models were compared using a likelihood-ratio statistic calibrated using 250 parametric bootstrap realisations. The Ashman separation statistic \citep{Ashman1994} and the minimum fitted component weight were also recorded. These quantities were retained as supporting diagnostics only and were not used to determine whether a branch was divided.

Accepted divisions were therefore treated as candidate line-of-sight overlaps rather than as definitive separate associations.

\section{Results}
The following sections present the results of the XMST validation programme. We first examine the recovery of the injected structures and the stability of the derived fracture scale. We then assess the effect of the published distance uncertainties on the recovered structures, evaluate the reddening diagnostic, compare the Percolation–Jenks procedure with the CDF and Mean Edge fracture criteria, and finally examine the sensitivity of the results to the adopted analysis thresholds.

\subsection{Monte Carlo results}
An initial set of 100 Monte Carlo realisations was used to assess the recovery performance of XMST and determine whether the principal validation statistics had converged. The experiment was subsequently extended to 1000 realisations. No material change was found in the recovery statistics between the two experiments, indicating that the initial 100 realisations were sufficient to characterise the overall behaviour of the algorithm.
Across the 1000 Monte Carlo realisations, all eight injected structures satisfied the adopted detection criterion in every case. Mean completeness was 0.9996, showing that almost all injected members were retained within their best-matching recovered structures. Mean purity was lower, at 0.7413. Under the stricter recovery criterion, requiring completeness of at least 0.8 and purity of at least 0.5, a mean of 6.08 of the eight injected structures was recovered per realisation. This distinction shows that high membership completeness does not necessarily mean that an injected structure has been recovered as a clean, distinct group.
The reduction in mean purity was dominated by the deliberately overlapping pair. These two templates had mean purities of 0.463 and 0.442 and remained merged in 985 of the 1000 realisations. In contrast, the six non-overlapping templates achieved mean purities between 0.809 and 0.860 while retaining essentially complete membership recovery. Their strict recovery rates ranged from 0.953 to 1.000. The lower overall purity therefore reflects primarily the deliberately difficult overlapping configuration rather than a uniform reduction in recovery quality across all injected structures.

\begin{figure}
    \centering
    \includegraphics[width=1\linewidth]{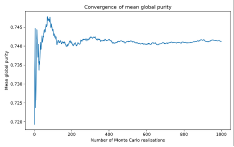}
    \caption{Convergence of the mean global purity across the 1000 Monte Carlo realisations.}
    \label{fig:placeholder}
\end{figure}

\subsubsection{Fracture-scale Stability }
The stability of the Percolation–Jenks fracture scale was assessed across the controlled Monte Carlo realisations. An initial set of 100 realisations was used to establish the behaviour of the method before the experiment was extended to 1000 realisations to test convergence. The derived fracture-scale distribution showed no material change as the number of realisations increased, indicating that the principal properties of the distribution had already converged within the initial experiment.

Across the 1000 realisations, the median percolation limit was approximately 29.69 pc. Application of Jenks Natural Breaks optimisation to the subcritical edge-length distribution produced a median final fracture scale of 17.028 pc, with a between-realisation standard deviation of 0.066 pc.

Bootstrap analysis produced a median within-realisation fracture-scale uncertainty of 0.076 pc. Both the between-realisation scatter and the bootstrap uncertainty were less than 0.5 per cent of the median fracture scale, demonstrating that the derived value was highly reproducible under changes in the positions and orientations of the injected structures and under resampling of the subcritical edge-length distribution.

The fracture-scale distribution contained a small lower-scale tail, with 110 of the 1000 realisations producing fracture scales below 16.9 pc. These realisations did not show a corresponding reduction in recovery performance. Their mean completeness was approximately 0.997 compared with approximately 1.000 for the remaining realisations, while mean purity remained approximately 0.741 in both subsets. The mean number of strict recoveries was also essentially unchanged. The lower-scale feature therefore does not appear to represent a distinct failure mode of the algorithm.

\begin{figure}
    \centering
    \includegraphics[width=1\linewidth]{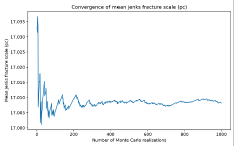}
    \caption{Convergence of the mean Jenks fracture scale across the 1000 Monte Carlo realisations, showing rapid stabilisation near 17.01 pc.}
    \label{fig:3}
\end{figure}

\begin{figure}
    \centering
    \includegraphics[width=1\linewidth]{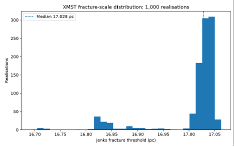}
    \caption{Distribution of the Jenks fracture scale across 1000 Monte Carlo realisations. The median fracture scale is 17.028 pc, with a small lower-scale tail.}
    \label{fig:4}
\end{figure}

\begin{figure}
    \centering
    \includegraphics[width=1\linewidth]{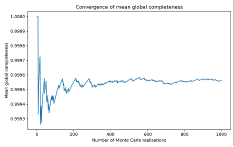}
    \caption{Convergence of the mean global completeness across the 1000 Monte Carlo realisations, stabilising at approximately 0.9996.}
    \label{fig:5}
\end{figure}

\subsection{Stability under Distance Uncertainty}

The effect of the published distance uncertainties was assessed using 100 uncertainty realisations of each of the first 100 Monte Carlo fields, giving a total of 10{,}000 additional spatial XMST reconstructions. For each realisation, stellar distances were perturbed using the asymmetric uncertainty treatment described in Section~2.6. The Cartesian coordinates, Delaunay graph, minimum spanning tree and Percolation--Jenks fracture scale were then recalculated independently.

The global fracture scale remained highly stable under these perturbations. Across the 10{,}000 uncertainty realisations, the median fracture scale was $17.251\,\mathrm{pc}$ with a standard deviation of $0.305\,\mathrm{pc}$, compared with a median of $17.028\,\mathrm{pc}$ for the corresponding unperturbed fields. The median change relative to the corresponding unperturbed field was $+0.245\,\mathrm{pc}$, while the median absolute change was $0.282\,\mathrm{pc}$. The typical change was therefore only approximately 1--2 per cent of the nominal fracture scale.

The memberships of individual recovered structures were more sensitive to the distance perturbations than the global fracture scale. Comparison of the nominal recovered groups with their best-matching perturbed counterparts gave a mean Jaccard Index of 0.433 and a median of the per-group median Jaccard values of 0.445. The median stellar membership persistence was 0.74, with approximately 44.7 per cent of stars having a persistence of at least 0.8. These statistics measure the stability of detailed group membership and should not be interpreted as the fraction of injected structures successfully detected.

The injected structures themselves remained readily detectable. Mean completeness decreased from 0.9994 in the unperturbed fields to 0.8605 following the distance perturbations, while mean purity decreased from 0.7445 to 0.7126. Despite these changes, 99.30 per cent of the injected-template cases continued to satisfy the adopted detection criterion of recovering at least 50 per cent of the injected membership. The published distance uncertainties therefore affected the detailed membership of the recovered structures considerably more strongly than their overall detection.

\subsection{Reddening Validation}
Initial testing investigated the inclusion of reddening directly within the XMST clustering space. This did not improve the recovery of the injected structures and the approach was therefore abandoned. Reddening was instead retained as a post-processing diagnostic, applied to candidate divisions within structures identified by the spatial XMST analysis.

The deliberately overlapping pair provided the principal test of this approach. Using spatial information alone, the two templates were recovered as separate structures in only 15 of the 1000 Monte Carlo realisations. Following application of the reddening diagnostic, the pair was recovered as two distinct structures in 446 of the 1000 realisations. Reddening therefore provided substantial additional information where the spatial MST alone was unable to resolve the superposition, although more than half of the deliberately overlapping cases remained unresolved. 

This improvement in separation was accompanied by a change in the overall purity–completeness balance. Before reddening-based subdivision, mean completeness and purity were 0.9996 and 0.7413, respectively. Following application of the reddening diagnostic, mean completeness decreased to 0.9467 while mean purity increased to 0.8038. The reddening stage therefore reduced contamination within the recovered structures, but at the cost of losing some genuine members. 

These results indicate that reddening can provide a useful secondary diagnostic for identifying possible line-of-sight superpositions, but does not provide a sufficiently reliable discriminator to form part of the primary XMST clustering metric. The spatial XMST solution was therefore retained as the primary detection stage, with reddening used to identify candidate subdivisions requiring further investigation.

\subsection{CDF and Mean Edge Comparison }
To compare the different fracture-scale criteria directly, the same 1000 Monte Carlo realisations were reanalysed using the CDF and Mean Edge methods in place of Percolation–Jenks. The CDF fracture scale was determined using the two-line fitting procedure described in Section 2.4.1, while the Mean Edge fracture scale was defined as the mean length of the global MST edges. All other aspects of the recovery analysis were unchanged.

Both alternative methods selected substantially larger fracture scales than Percolation–Jenks. The median CDF fracture scale was 34.728 pc and the median Mean Edge scale was 24.990 pc, compared with 17.028 pc for Percolation–Jenks. 

The larger fracture scales were accompanied by substantially lower purity. Mean purity decreased from 0.7413 with Percolation–Jenks to 0.4574 using the Mean Edge criterion and 0.0779 using the CDF criterion. The mean number of injected structures satisfying the strict recovery criterion was 6.08 of eight for Percolation–Jenks, compared with approximately 3.0 for the Mean Edge method and 0.5 for the CDF method. Neither alternative criterion spatially resolved the deliberately overlapping pair in any of the 1000 realisations, compared with 15 successful separations using Percolation–Jenks.

\begin{table*}
	\centering
	\caption{Comparison of MST fracture methods.}
	\label{tab:fracture_methods}
	\begin{tabular*}{\textwidth}{@{\extracolsep{\fill}}lcccc}
		\hline
		Fracture method & Median scale & Mean strict recoveries & Purity & Overlap pair \\
		 & (pc) & (of 8) & & resolved \\
		\hline
		Percolation--Jenks & 17.028 & 6.08 & 0.7413 & 15/1000 \\
		Mean Edge          & 24.990 & 3.0  & 0.4574 & 0/1000  \\
		CDF                & 34.728 & 0.5  & 0.0779 & 0/1000  \\
		\hline
	\end{tabular*}
\end{table*}

\subsection{Sensitivity Tests}
The recovery statistics were insensitive to changes in the adopted minimum retained group size over the range tested. For $N_{\mathrm{min}} = 10$, $15$ and $20$, mean completeness, purity, detection rate and strict-recovery statistics were unchanged because all best-matching recovered structures contained at least 26 stars. Lowering $N_{\mathrm{min}}$ to 5 likewise did not alter the recovery statistics of the injected structures, although the lower threshold could permit additional small components to be retained.

The detection statistic was also insensitive to the adopted membership threshold. All eight injected structures remained detected when the required recovered fraction was varied between 0.4 and 0.7. Varying the strict completeness threshold between 0.7 and 0.9 produced no material change because the recovered memberships were already close to complete.

The strict-recovery statistic was more sensitive to the adopted purity threshold. For purity thresholds of 0.4, 0.5, 0.6 and 0.7, the mean numbers of strict recoveries were 7.78, 6.08, 5.75 and 5.29 structures out of eight, respectively. This reflects the expected effect of progressively requiring the recovered structures to contain fewer contaminating members.

The adopted thresholds therefore affect the numerical value of the strict-recovery statistic, particularly through the purity requirement, but do not alter the overall result that the injected structures are consistently detected with very high completeness.

\section{Discussion}
We have presented XMST, an extended minimum spanning tree methodology for the identification of stellar structures within complex astronomical datasets. The method uses a combination of percolation analysis and Jenks Natural Breaks optimisation to determine an objective fracture scale directly from the global MST.

\subsection{Why Percolation--Jenks works}

The principal advantage of the Percolation--Jenks procedure is that the two stages address different parts of the fracture-scale problem. The percolation stage identifies the transition at which progressively longer MST edges begin to connect previously separate local structures into a substantially larger network. In XMST this is defined by the largest increase in the membership of the largest connected component, providing an upper limit on the edge-length range considered when determining the final fracture scale. Jenks Natural Breaks optimisation is then applied only to the edge lengths below this limit, separating the shorter-edge population from the longer connections remaining within the subcritical distribution.

Restricting the Jenks analysis in this way reduces the influence of long inter-structure and field-spanning edges within the upper tail of the global MST edge-length distribution. The resulting fracture scale is therefore determined primarily from the edge-length regime associated with locally connected structure rather than from the complete distribution. The Monte Carlo experiments show that, for the fields considered here, this produces a highly reproducible fracture scale of approximately $17\,\mathrm{pc}$ while retaining almost complete recovery of the injected members.

The procedure should not be interpreted as identifying a unique physical scale of star formation. Rather, it provides an objective operational scale at which a particular MST can be fractured into candidate stellar structures. Its usefulness is therefore established by the stability of the derived scale and the recovery properties of the resulting subtrees, rather than by assigning direct physical significance to the numerical value of the fracture scale itself. In this context, objective refers to the reproducible, data-driven selection of the fracture scale once the analysis procedure and its parameters have been specified; it does not imply that XMST is parameter-free.

\subsection{Comparison with CDF and Mean Edge Fracturing}

Both the CDF and Mean Edge criteria selected substantially larger fracture scales than the Percolation--Jenks procedure in the controlled Monte Carlo experiments. The median fracture scales were $34.728\,\mathrm{pc}$ for the CDF method and $24.990\,\mathrm{pc}$ for the Mean Edge method, compared with $17.028\,\mathrm{pc}$ for Percolation--Jenks. The larger thresholds allowed most injected members to remain connected, but also retained substantially more field stars and connections between neighbouring structures. Consequently, mean purity decreased from 0.7413 with Percolation--Jenks to 0.4574 using the Mean Edge criterion and 0.0779 using the CDF criterion.

This behaviour appears to arise from the use of the complete MST edge-length distribution by both alternative estimators. Long edges connecting locally coherent structures to each other and to the surrounding field contribute directly to the derived threshold. In the complex multi-structure fields considered here, this shifts the selected fracture scale towards longer edge lengths. The resulting subtrees can therefore retain very high membership completeness while failing to recover the injected structures as clean, distinct groups.

Percolation--Jenks reduces this effect by first restricting the edge-length distribution to the subcritical regime identified by the percolation stage before determining the final fracture scale. The comparison therefore indicates that, for the fields investigated here, Percolation--Jenks provides a substantially better balance between retaining genuine association members and rejecting unrelated stars than either the CDF or Mean Edge criteria.

\subsection{The fracture scale is statistically stable}

The Percolation–Jenks fracture scale was highly reproducible across the controlled Monte Carlo realisations. Across the 1000 unperturbed realisations, the median fracture scale was 17.028 pc, with a between-realisation standard deviation of 0.066 pc, while the median within-realisation bootstrap uncertainty was 0.076 pc. The similarity of these values indicates that the variation produced by changing the positions and orientations of the injected structures was comparable to that obtained by resampling the subcritical MST edge-length distribution.

The fracture scale also remained stable when the published distance uncertainties were propagated through the spatial clustering procedure. Across the 10,000 uncertainty-perturbed reconstructions, the median fracture scale was 17.251 pc, with a median change of +0.245 pc relative to the corresponding unperturbed field and a median absolute change of 0.282 pc. Although the distance perturbations increased the scatter in the derived scale, the typical change remained only approximately 1–2 per cent of the nominal fracture scale.

The small lower-scale tail found in the unperturbed Monte Carlo distribution also produced no corresponding deterioration in recovery performance. Realisations with fracture scales below 16.9 pc retained essentially the same purity and strict-recovery statistics as the remainder of the sample. Taken together, these tests indicate that the Percolation–Jenks scale is not strongly dependent on a particular Monte Carlo configuration, modest variations in the MST edge-length distribution, or the published stellar-distance uncertainties.

\subsection{Recovery, purity and the limits of a global threshold}

The Monte Carlo experiments show that XMST recovers the injected structures with very high completeness, while the purity of the recovered subtrees is more variable. Across the 1000 unperturbed realisations, mean completeness was 0.9996 and mean purity was 0.7413. Much of the reduction in purity was associated with the deliberately overlapping pair, while the six non-overlapping templates retained essentially complete membership recovery with substantially higher purities.

This distinction is important because recovering the members of an injected association does not necessarily mean that the association has been recovered as a clean, isolated structure. A subtree may contain nearly all genuine members while also retaining field stars or members of a neighbouring structure connected by edges shorter than the adopted fracture scale. High completeness therefore demonstrates successful detection of the underlying structure, but does not by itself establish a clean membership boundary.

The distance-uncertainty analysis reinforces this interpretation. Following perturbation of the stellar distances, 99.30 per cent of injected-template cases continued to satisfy the adopted detection criterion, while mean completeness decreased to 0.8605 and mean purity to 0.7126. The changing memberships were concentrated preferentially towards spatially peripheral and locally sparse regions, while the structural cores remained substantially more persistent. Distance uncertainty therefore affects the detailed boundaries of the recovered structures more strongly than their overall detection.

Part of this behaviour results from the use of a single global fracture scale across a field containing structures with different internal densities and morphologies. A threshold that preserves the diffuse extensions of one association may also retain short connections to unrelated stars elsewhere in the field. Conversely, reducing the threshold to improve purity risks fragmenting genuine extended structure. The global fracture scale should therefore be regarded as a compromise between completeness and contamination rather than as an exact physical boundary applicable to every association.

For this reason, XMST subtrees should be interpreted as candidate stellar associations rather than definitive membership catalogues. Membership-persistence statistics provide a means of identifying the stars whose assignments are most robust to the published distance uncertainties, while additional astrophysical information, including kinematics, photometry, spectroscopy and age indicators, may be required to refine individual memberships and to separate physically distinct populations that remain connected spatially.

\subsection{Reddening as a secondary diagnostic}

Reddening improved the separation of the deliberately overlapping structures, increasing the number of successfully separated pairs from 15 to 446 out of the 1000 Monte Carlo realisations. This demonstrates that reddening can provide useful additional information where spatial data alone are unable to distinguish between two populations. However, more than half of the overlapping pairs remained unresolved, so reddening cannot be regarded as a reliable solution to line-of-sight overlap on its own.

There was also a clear trade-off between purity and completeness. Following the reddening analysis, mean purity increased from 0.7413 to 0.8038, while mean completeness decreased from 0.9996 to 0.9467. Reddening therefore helped to remove contamination and separate some mixed structures, but in some cases this also resulted in genuine members being removed or an injected structure being divided.

The effectiveness of reddening as a discriminator is likely to depend on the separation between the extinction distributions of the populations relative to their internal scatter and observational uncertainty. Where the distributions overlap strongly, reddening provides relatively little additional information. Where they are more clearly separated, it can provide useful evidence that an apparently single spatial structure contains more than one population.

Attempts to include reddening directly within the MST as an additional clustering dimension did not improve the recovery of the injected structures and were therefore abandoned. The results instead support its use as a post-processing diagnostic after the spatial clustering has been completed. Any subdivision identified in this way should consequently be treated as evidence for a possible line-of-sight overlap rather than as a definitive association boundary.

Other astrophysical information, particularly kinematics, may provide a more effective means of resolving such cases. Proper motions and radial velocities could distinguish populations that overlap spatially but have different motions. These tests are outside the scope of the present work and are left for subsequent analysis.

\subsection{Global versus local fracture scales}
Throughout this work, a single global fracture scale has been applied across the complete stellar field. This has the advantage of providing an objective and reproducible segmentation criterion, and the Monte Carlo experiments show that the resulting Percolation–Jenks scale is highly stable. However, a single global value cannot account fully for local variations in stellar density or for differences in the internal scales and morphologies of individual associations.

This limitation is likely to be most important in fields containing a mixture of compact clusters, diffuse associations and hierarchical structure. A fracture scale that is appropriate for one region may be too large elsewhere, retaining unrelated field stars or joining neighbouring structures. Conversely, adopting a smaller global fracture scale to improve purity in denser regions may fragment genuine diffuse structures elsewhere.

The results presented here therefore support the use of a global Percolation–Jenks scale as a robust first-order segmentation of the field, while also demonstrating the limitations of imposing a single threshold across structures with different local properties. One possible extension would be to determine fracture scales locally, for example by analysing overlapping spatial regions independently and subsequently combining the recovered structures. Such an approach could allow the clustering scale to respond to local structure while retaining the same objective Percolation–Jenks procedure.

A locally adaptive implementation has not been tested in the present work and would require a separate validation programme. In particular, it would be necessary to determine how structures recovered in neighbouring regions should be reconciled and whether locally varying fracture scales introduce additional fragmentation or boundary effects. This is therefore left for future work.

\subsection{Scope of the validation}

The empirical templates used in the Monte Carlo experiments were selected from structures recovered in the nominal XMST segmentation of the Q25 catalogue and therefore do not provide an independent external validation of astrophysical membership. Instead, the experiments were designed to test the behaviour of the method using realistic, non-parametric stellar morphologies while controlling their position, orientation, degree of spatial overlap and distance uncertainty.

The results therefore demonstrate the recovery and stability of XMST for the range of structures represented by these templates, rather than establishing that the method will recover every possible association morphology or density distribution. An independent comparison with published association memberships and alternative clustering methodologies is required to assess that broader question and is reserved for the subsequent application paper.

\section{Conclusions}

We have presented XMST, an extended minimum spanning tree framework designed to identify spatially coherent stellar structures while providing an objective procedure for determining the fracture scale. XMST combines percolation analysis with Jenks Natural Breaks optimisation, using the percolation stage to restrict the edge-length distribution before the final fracture scale is determined. The resulting procedure avoids the need to select the MST fracture scale by eye or from an assumed functional form.

Tests using 1000 controlled Monte Carlo realisations showed that the Percolation–Jenks fracture scale was highly reproducible, with a median value of 17.028 pc and a between-realisation standard deviation of 0.066 pc. The injected structures were recovered with a mean completeness of 0.9996 and mean purity of 0.7413. In the same tests, both the CDF and Mean Edge criteria selected substantially larger fracture scales and produced poorer recovery of the injected structures as clean, distinct groups.

Propagating the published distance uncertainties through 10,000 additional XMST reconstructions produced only a small change in the characteristic fracture scale, with a median absolute shift of 0.282 pc. The injected structures continued to satisfy the detection criterion in 99.30 per cent of cases. Individual membership was less stable, however, with the least persistent members preferentially located towards the spatially peripheral and locally sparse parts of the structures. This suggests that XMST is more robust in identifying the existence and cores of stellar structures than in defining their exact outer membership.

Significant spatial overlap remains a limitation. The deliberately overlapping pair was separated using spatial information alone in only 15 of the 1000 realisations. Application of the reddening diagnostic increased the number of resolved cases to 446 and improved mean purity from 0.7413 to 0.8038, but at the cost of reducing completeness from 0.9996 to 0.9467. Reddening is therefore useful as a secondary diagnostic, but is not sufficiently reliable to resolve overlapping populations on its own.

XMST should therefore be regarded as a method for identifying robust candidate stellar structures rather than as a means of assigning definitive astrophysical membership. The present work establishes the behaviour and stability of the spatial method under controlled conditions. Detailed application to the Gaia OB-star catalogue, comparison with existing clustering methods, and the inclusion of additional information such as kinematics are left to subsequent work.

\section*{Acknowledgements}

This work has made use of data from the European Space Agency (ESA) mission Gaia (https://www.cosmos.esa.int/gaia), processed by the Gaia Data Processing and Analysis Consortium (DPAC). Funding for the DPAC has been provided by national institutions, in particular the institutions participating in the Gaia Multilateral Agreement.
The author also thanks the authors of the Quintana et al. catalogue for making their data available.
Generative AI tools were used during manuscript preparation for language editing and assistance with code development. All scientific decisions, analysis, code execution, verification and interpretation of the results were carried out by the author.

\section*{Data Availability}

The stellar catalogue used in this work is publicly available through the VizieR catalogue service as catalogue J/MNRAS/538/1367 (Quintana et al. 2025). The XMST source code, configuration files and compact validation outputs used in this work are available at the XMST Stage 1 GitHub repository, https://github.com/mjgallawayastro-stack/XMST-Stage-1. The complete Monte Carlo validation outputs and distance-uncertainty analysis associated with this work are archived on Zenodo under DOI 10.5281/zenodo.221126



\bibliographystyle{rasti}
\bibliography{example} 





\bsp	
\label{lastpage}
\end{document}